Complex Adaptive Systems Modeling

 

# Game theory models for communication between agents: a review

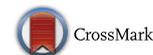

Aisha D. Farooqui[1] and Muaz A. Niazi[2]*

*Correspondence:
muaz.niazi@ieee.org
[2] Computer Science
Department, COMSATS
Institute of IT, Islamabad,
Pakistan
Full list of author information
is available at the end of the
article

## Abstract

In the real world, agents or entities are in a continuous state of interactions. These interactions lead to various types of complexity dynamics. One key difficulty in the study of complex agent interactions is the difficulty of modeling agent communication on the basis of rewards. Game theory offers a perspective of analysis and modeling these interactions. Previously, while a large amount of literature is available on game theory, most of it is from specific domains and does not cater for the concepts from an agent-based perspective. Here in this paper, we present a comprehensive multidisciplinary state-of-the-art review and taxonomy of game theory models of complex interactions between agents.

**Keywords:** Game theory applications, Agent-based approach, Complex adaptive systems , Complex systems modeling , Complex networks, Economics

## Background

In the real world, agents or entities are in a continuous state of interactions (Niazi et al. 2011). Examples of these include the continuously interacting agents in the stock market (Bonabeau 2002). These agents and systems can be adaptive in nature and can also evolve. Their current behavior can depend on the past so they often learn from history.

The interaction of agents leads to a wide variety of complexity dynamics (McDaniel and Driebe 2001). Complexity arises due to non-linear agent interactions. The behavior of such non-linear systems can be chaotic and unpredictable. Complex adaptive systems (CAS) in the natural world (Niazi et al. 2011) and complex physical systems (CPS) (Winsberg 2001) in man-made systems are examples of such agent interactions.

One key difficulty faced by Complexity researchers is in the modeling of communication and complex agent interaction (Niazi and Hussain 2012). Modern communication systems are often composed of hierarchical complex systems. These systems can be modeled as multiagent systems using agent-based modeling (ABM). Modeling CAS and CPS using ABM not only allows for prediction of outcomes but also helps in terms of gaining an understanding of the complex inter-connnections and interactions (Epstein 2008). However, a key issue in such models is to understand the dynamics of agent interaction. Game Theory offers techniques and tools for modeling communication problems among agents.

Game theory offers a perspective of analysis and modeling of these interactions (Carmichael 2005). It is a discipline that studies decision making of interactive entities (Dixit and Skeath 1999). We can say that strategic thinking is  perhaps the most recognized essence of game theory.

Previously, while a large amount of literature is available on game theory, most of it is  focused on specific domains like Biology, Economics, and Computer Science (Shoham and Leyton-Brown 2008). Game theory has also been used in business to model interactions of stakeholders etc.

To the best of our knowledge, there is an absence of a state-of-the-art reviews  of game theoretical literature from the agent-based modeling perspective. This paper presents a comprehensive review of game theory models and their applications. Additionally, a taxonomy of classes of games is also presented.

The paper is organized as follows: first, we give an overview of game theory and present a taxonomy of games. This is followed by literature review in the next section. Then, in the discussion we classify games and discuss open problems before concluding the paper.

## Game theory overview

While the essence of game theory has perhaps practically applied itself since life presented itself on this planet, formal literature on the topic can be traced back to the work of Von Neumann and Morgenstern (1944). They worked on zero-sum games. Then in the 1950s, Nash's work resulted in significant advancement of this field (Nash 1950). Subsequently, Game theory has  since been used in many different fields like biology (Hofbauer and Sigmund 1998), politics and other domains (Morrow 1994).

Game theory presents a technical analysis of strategic interactions (Shoham and Leyton-Brown 2008). These strategic interactions are concerned with the interaction of decision makers in the game (Geckil and Anderson 2009). The behavior of a decision maker in game theory models is called "strategic" and the action performed while making any move is called a "strategy". Strategy considers how agents act, what they prefer, how they make their decisions,  and their behaviors etc. These interactions can be complex as the action of even a single  agent can influence other agents and vice versa. Game theory can thus be considered as a powerful tool to model and understand complex interactions.

One way of classifying game theory models is to divide them into cooperative and non-cooperative games (Shoham and Leyton-Brown 2008). In cooperative games, we focus  on a set of agents. Whereas, in non-cooperative games the focus is on the development of models of  interactions, preferences, and so on,   with a focus on individual agents.[1] It can model different types of games including zero-sum (Shoham and Leyton-Brown 2008), stochastic (Mertens and Neyman 1981), repeated Aumann and Maschler (game of fairness as if player cuts unequ), Bayesian (Böge and Eisele 1979) and congestion (Rosenthal 1973).

---

[1] Literature usually considers cooperative and non-cooperative as conflicting and non-conflicting game theory. But we are following definition of Shoham et al. that cooperative game theory focus on modeling set of players and non-cooperative models individual player (Shoham and Leyton-Brown 2008).

**Multidisciplinary nature of game theory**

Game theory can be seen everywhere in living systems, in general, and human society, in particular. In personal life as well as in professional life, every day we are faced with decisions which often can be simplified using game theory. There are different areas where game theory has been applied such as Economics, Politics etc.

(Shoham and Leyton-Brown 2008). Algorithmic game theory is an example of application in computer science (Roughgarden 2010). Biologists have used it to learn species behaviors (Hofbauer and Sigmund 1998). In mathematics, there is a complete branch that studies decision-making process (Mazalov 2014). It also has its influences in business (Geckil and Anderson 2009). It can model interactions of stakeholders, dynamics in interest rates etc.

Dixit and Skeath (1999) note that we can use game theory mainly in three ways that are an explanation, prediction, and prescription.

*Explanation*

Game theory can be used to explain insights of a situation like why that happened, what were the causes, Effects of that happening etc. We can do a complete case study by using game theory.

*Prediction*

Game theory studies decision makers (autonomous agents) that have actions to take, preferences that what they want, different options which they can choose etc. By analyzing these actions, preferences, options etc we can predict different moves of agents on different types of situation.

*Prescription*

If we can analyze agent actions, strategies etc to predict its moves, then we can definitely give advice about different moves to agents. It means we can provide a sophisticated model for future decision-makings.

Now let us consider basic concepts of game theory.

**Basic concepts**

Dixit and Nalebuff (1993) have defined Game theory as:

**Definition 1**   The branch of social science that studies strategic decision-making.

Another definition is by Hutton (1996):

**Definition 2**   An intellectual framework for examining what various parties to a decision should do given their possession of inadequate information and different objectives.

*Shoham and Leyton-Brown (2008) have defined game theory as:*

**Definition 3**   Game theory is the mathematical study of interaction among independent, self-interested agents.

In the oxford dictionary, self-interested means self-seeking or self-serving. Anyone who is self-interested is concerned strongly with own interests. This seems selfishness of someone who do not consider others interests.

However, in game theory, these are actually intelligent agents and their behavior is based on artificial intelligence models (Wooldridge 2009). These are autonomous entities, with their own description of world states and they behave accordingly (Shoham and Leyton-Brown 2008). Unfortunately, there is no universal definition of the agent but autonomy is one of the basic properties of the agent.

In Computer Science, Algorithmic game theory is used (Roughgarden 2010). It combines game theory together with computer science. It focuses on creating algorithms for strategic interactions, calculating Nash equilibrium etc.

### Game

Carmichael (2005) has defined games as:

**Definition 4** A scenario or situation where for two or more individuals, their choice of action or behavior has an impact on the other (or others).

The game consists of several things such as

- players
- strategies (actions taken while interactions)
- payoffs (utilities gained)
- payoff function (calculates utility against each strategy)
- and of course, game rules.

Geckil and Anderson (2009) has defined game as:

**Definition 5** A game-theoretic model is an environment where each decision maker's actions interact with those of others

### Game representation

There are mainly two ways to represent the game. Normal-form is simply a matrix that describes strategies and payoffs of the games (Morrow 1994). Another representation is extensive-form, which is a tree-like structure (Morrow 1994). Extensive-form contains more information than normal-form like a sequence of player moves. However, there are games that require richer representation such as infinite repeated games. To represent such games we have Beyond Normal-Extensive form (Shoham and Leyton-Brown 2008).

### Decision theorem

Game theory has two decision theorems known as maximin and minimax (Mazalov 2014). The minimax theorem minimizes the loss of a player. The maximin theorem used to maximize the benefit gain by the player.

## Games taxonomy

We saw different types of games in the literature review. These games were presented using three types of game representations. Normal-form, extensive-form and beyond normal and extensive-form games (Shoham and Leyton-Brown 2008). We proposed a taxonomy of games based on these three game representation types. See Fig. 1.

The taxonomy mainly classifies games into three types, as there are three types of representations. Then it further classifies games that are included in both normal-form games and extensive-form games. Games included in both because a normal-form representation can be derived from extensive-form games. Beyond normal and extensive form includes those games that need richer representation. These games can be infinite and undetermined. Therefore, that it is difficult to represent them in first two representations.

These games have been discussed in literature according to game representation types but is not presented as the taxonomy in this paper demonstrates. There are previously given taxonomies, but these are specific to the two-player game. Kilgour and Fraser have presented a taxonomy discussing ordinal games (Kilgour and Fraser 1988). Rapoport and Guyer (1978) have presented another taxonomy considering $2 \times 2$ games. The taxonomy given in this paper is not specific and is based on the type of game representation.

### *Normal-form games*

It is conceptually straightforward strategic representation (Morrow 1994). It describes all observable and possible strategies and the utility against each strategy. It can represent all finite games and taken as a universal representation of games. It uses a matrix to represent strategic interactions of players in a matrix form. It consists of

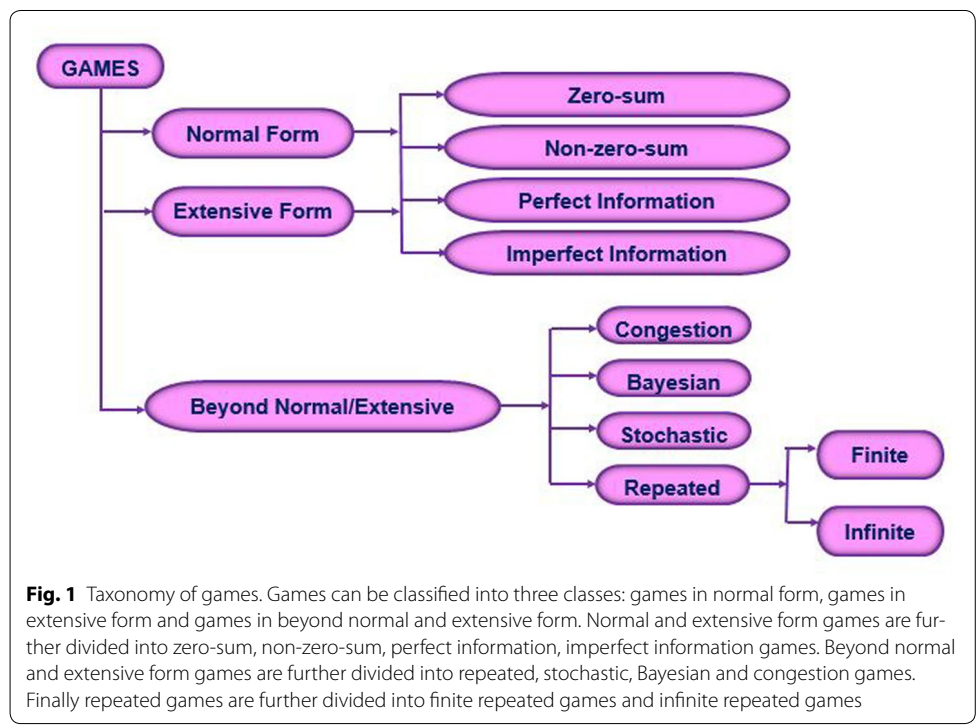

**Fig. 1** Taxonomy of games. Games can be classified into three classes: games in normal form, games in extensive form and games in beyond normal and extensive form. Normal and extensive form games are further divided into zero-sum, non-zero-sum, perfect information, imperfect information games. Beyond normal and extensive form games are further divided into repeated, stochastic, Bayesian and congestion games. Finally repeated games are further divided into finite repeated games and infinite repeated games

- Set of players
- Strategy space, a set of all strategies of a player
- Payoff function, it calculates the utility against each strategy.

Table 1 adapted from Morrow (1994) shows a normal form representation of Matching Pennies game. If both P1 and P2 get heads, P1 will take both coins else P2 will win and take both coins. The numbers 1 and −1 shows the utility gained or loosed by players.

### Extensive-form games

It is an alternative way of representing games in a tree-like structure. It defines different stages of the game. Moves, choices, and actions defined according to each stage. We can derive a normal-form representation from extensive representation. Morrow (1994) described Matching pennies game in extensive form representation. See Fig. 2.

### Beyond normal/extensive games

There are games needs richer representation like repeated games (Shoham and Leyton-Brown 2008). These can be finite or infinite. Therefore, that it is difficult to represent them in normal/extensive forms. The games included here are.

- Repeated games: These are also called stage games. Players play these games multiple times (Aumann and Maschler 1995).
- Stochastic games: These are also called Markov games. There are stages in the game. Every stage represents the state of a game from a finite set of game states. The player has a set of actions that consists of many finite actions (Mertens and Neyman 1981).
- Bayesian games: These are games of incomplete information. Players select their strategies according to Bayes' Rule (Böge and Eisele 1979).
- Congestion games: These games are the class of non-conflicting games (Rosenthal 1973). In these games, all the players have same strategy set. The result of every player relies upon the strategy it picks and all other players picking the same strategy.

## Complex adaptive systems

Complex systems have special types of systems known as Complex adaptive systems (Mitchell 2009). These systems have the dynamic environment and non-linear interaction of components. The amazing thing for researchers is that these systems are composed of so simple components and exhibits emergent behavior when combined together. Such systems can be understood only by considering all components collectively.

**Table 1 Matching pennies: game in normal-form (This table is adapted from Morrow (1994))**

| P1 | P2 | |
|---|---|---|
| | H | T |
| H | (1, −1) | (−1, 1) |
| T | (−1, 1) | (1, −1) |

In this game if both players show same side of coins that is both shows head or tail then P1 wins and P2 looses both coins. If both players shows different sides then P2 wins and P1 looses both coins

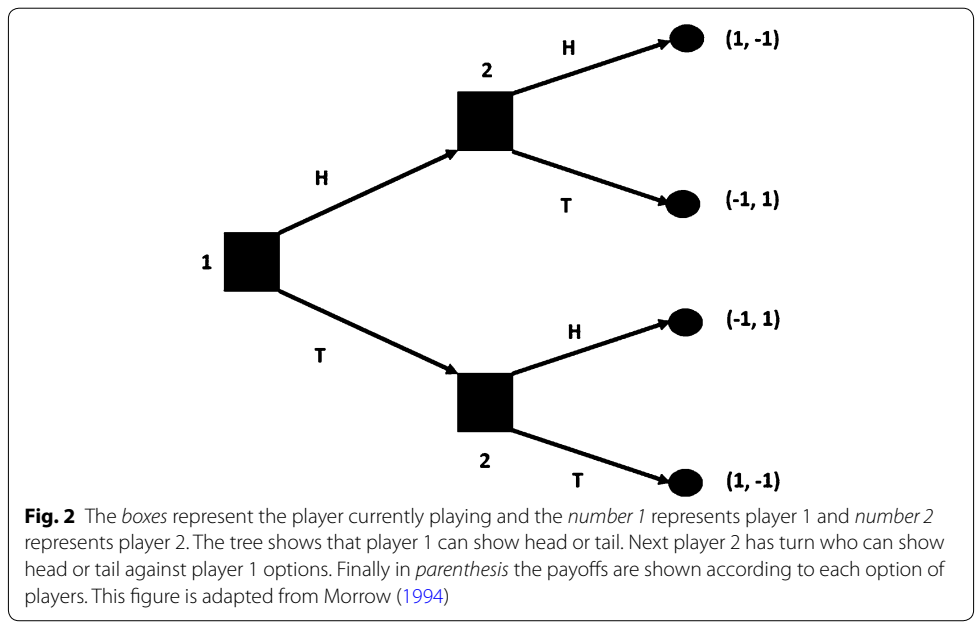

**Fig. 2** The *boxes* represent the player currently playing and the *number 1* represents player 1 and *number 2* represents player 2. The tree shows that player 1 can show head or tail. Next player 2 has turn who can show head or tail against player 1 options. Finally in *parenthesis* the payoffs are shown according to each option of players. This figure is adapted from Morrow (1994)

### Nonlinear agent interaction

Complex adaptive systems are subset of dynamic non-linear system (McDaniel and Driebe 2001). In non-linear agent interactions, the inputs are inversely proportional to output (Lansing 2003). In these amazing systems, small changes can results in a big change and vice versa. Mathematically, the behavior of the non-linear system can be described as non-linear polynomial equations.

There can be more than one attracters in non-linear systems (Socolar 2006). These attractors are of different types with complicated limit cycles. The trajectories are restricted to areas that have unstable limit cycles.

### Agent-based computing

Agent-based computing is a wide domain (Niazi and Hussain 2011). The agent here can simply a software providing any service. Or it can be fully autonomous agent whose behavior based on artificial intelligence. Agent-based computing should not be confused with other terms in artificial intelligence. Such terms are agent-oriented programming, multi-agent oriented programming, and agent-based modeling. These all are actually collected together in agent-based computing.

Now in the next section, we will present a review on available game theoretic literature.

## Review

In the previous section, we gave an overview of game theory and presented a taxonomy of games. In this section, we will explore available game theoretic literature.

### Zero-sum game theoretic models

Zero-sum games are the mathematical representation of conflicting situations (Washburn 2003). In these games, the total of gains and losses is equal to zero. Application of these game theoretic models can be seen in different fields like network security (Perea and Puerto 2013) and resource allocation (Zhou et al. 2011). There are also different types of games. Such as zero-sum games with incomplete information and large Zero-sum games.

Al-Tamimi et al. (2007) have discussed Q-learning designs for the zero-sum game. By using a model-free approach they obtained a solution for the game. Autopilot design for the F-16 plane is performed that shows productiveness of method.

Daskalakis et al. (2015) have proposed no-regret algorithm. This zero-sum game theoretic algorithm achieves regret when applying against an adversary. After using the algorithm, quadratic improvement can be identified on convergence rate to game value. The lower bound for all distributed dynamics is optimal. This happens when payoff matrix information is unknown to both players. But if they know they can compute minimax strategies privately.

Bopardikar et al. (2013) have studied larger zero-sum games. In these games, players have a large number of options. It proposes two algorithms. The Sampled Security Policy algorithm is to compute optimal policies. Then Sampled Security Value algorithm computes the level of confidence on the given policy.

Moulin and Vial (1978) have proposed a class of games called strategically zero-sum games. These games have special payoff structure. The mixed equilibrium of these games cannot be improved. The properties of games via a large body of correlation scheme is also described.

Sorin (2011) have worked on repeated zero-sum games. They described current advancement in these games especially together with differential games. They first define models of repeated games and differential games. Then they discuss issues related to these models.

Seo and Lee (2007) have considered conflicting zero-sum game that involves decision-making process. This is an experimental study on trained monkeys. Monkeys take binary choices in the computer-simulated conflicting game. The study described the decision-making process adaptive in both human and animals.

Zoroa et al. (2012) have modeled a perimeter patrol problem. They used Zero-sum discrete search games as a framework for their study. They studied problem occurred in cylindrical surface. The problem in the linear set having cyclic order is also studied. Optimal strategies are found via computer code.

Xu and Mizukami (1994) have studied systems of state space. They obtained saddle-point by a constructive method. It describes that there can be several saddle-point solutions for the system. When several saddle-points exist, this universal system differs from the state space system. They found possible conditions for the existence of saddle-point.

Ponssard and Sorin (1980) have discussed zero-sum games with incomplete information. They discussed two ways to determine information of states. It can be obtained via independent chance moves or the unique one. Unique moves cause dependence in state information. Thus, it is complicated to analyze. Several results acquired in the independent case have their equivalent in dependent one.

Chen and Larbani (2006) have proposed undetermined utility matrix game. They worked for the solution of decision-making problem (MADM). This decision making deals with prioritization of alternatives considering several attributes. Here weights of an MADM problem obtained with a fuzzy decision matrix. Finally, equilibrium solution is also obtained.

Li and Cruz (2009) have studied deception. They used a zero-sum game model with an asymmetrical structure. This paper considers the relationship between information and decision-making to understand deception. In these games, the first player gets extra information. Whereas the second player has the power to inject deception. The paper also classifies deception into active deception and passive deception.

Ponssard (1975) have worked on the zero-sum game in the normal form. They described that these games are equal to a linear program (LP). In these games, the player's behavioral strategies are represented in variables. In normal form game variables are used to represent the player's mixed strategies.

Wang and Chen (2013) have obtained feedback saddle-point for the zero-sum differential game. The game is between counter-terror measure and economic growth. It uses Hamilton-Jacobi-Isaac's equation to obtain saddle-point. The saddle-point obtained, strengthens the government counter-terror and weakens the terrorist organizations.

Van Zandt and Zhang (2011) have studied equilibrium value for Bayesian zero-sum games. The conditions are characterized for equilibrium value and strategies. These games have a parameter to obtain payoff function and strategies for every player. The information of every player is modeled as a sub-$\sigma$-field to obtain optimal strategies.

Marlow and Peart (2014) have studied soil acidification. They described a zero-sum game between a sugar maple and American beech. The negative impact of soil acidification on sugar maple supports beech in the game. The model lay down the findings of this study and other evidence of soil acidification. The results suggest re-examining the cost-effectiveness of chemical remediation.

### *2-player zero-sum games*

Mertens and Zamir (1971) have also discussed the two-person zero-sum game with incomplete information. These games are studied in a repetitive form. As a result, the game value is obtained with n repetitions. This is previously discussed by Harsanyi. However, still this paper is completely independent on its own.

Chang and Marcus (2003) have studied two-person zero-sum game. They considered optimal equilibrium game value and then analyzed error bounds. After that, they discussed methods that calculate the value of subgame.

Méndez-Naya (1996) have discussed 2-players continuous games. These games have set of pure strategies. These games also have right-sided semi-open real intervals and continuous payoff functions. The paper described conditions for game value in the mixed game. It is proved that there is no assurance that mixed extended Zero-sum game has a value but there can be a value.

Qing-Lai et al. (2009) have proposed an algorithm for 2-D systems. It solves two-players zero-sum games. It obtains saddle-point by using adaptive critic technique. The optimal control policies have been computed using neural networks. The algorithm can be implemented without system model.

Zhang et al. (2011) have proposed an iterative algorithm. It obtains optimal solutions for the non-affine nonlinear zero-sum game. This is a two-player game with quadratic performance index. One player minimizes the performance index while other maximizes it. This study held to facilitate this minimax problem. The optimal strategy has obtained an order of state trajectories and Riccati differential equations. Finally, the simulation shows successful results of this iterative method.

Gensbittel (2014) has worked on zero-sum incomplete information games. The author extended the CAV (U) Theorem of Aumaan–Maschler (Aumann et al. 1995). In this paper, the presented results are for infinite repeated games. Finally, the paper provides optimal strategies for players in 2-players game having length n.

Bettiol et al. (2006) have considered Zero-sum state constrained differential games. The study proves bolza problem for two-player differential games. It shows that lower semi-continuous value function exists in differential games. The optimal strategy is created and the value function is characterized by viscosity solutions.

### Beyond 2-player zero-sum games

Initially, the zero-sum game is a 2-player game (Von Neumann and Morgenstern 1953). In which one player has to win and other has to loose the game. The following papers show that researchers have worked on beyond 2-player game.

Moulin (1976) has worked on beyond 2-player Zero-sum games. First, this study describes a large family of abstract extension. Then these extensions are classified based on information exchanged. Finally, characterization of all possible values gained from this abstract extension is described.

Okamura et al. (1984) have studied three-player zero-sum games. They investigated the learning of the behavior of variable-structure stochastic automata in a game. These automata have learning capabilities and can update their actions. The players have a lack of information of payoff matrix. After every play, the environment, responds to automaton actions. After this, players update their strategies.

### Decision theorems

Sauder and Geraniotis (1994) have worked on maximin and minimax theorems. They formulated signal detection process as two-players zero-sum game. The two-players are the detector designer and the signal designer. The signal detection problem arises when analyzing the signal is genuine or deceptive. Finally, results are validated via simulation.

Hellman (2013) have focused on rational belief system. The study got the basis from the work of Aumann and Dreze. They described that players have common knowledge of rationality. Whereas in this article, it is argued that there is no need of common rationality. Finally, it is shown that the expected payoff in the game is only the minimax value.

Ponssard (1976) have discussed minimax strategies. These are prohibited to give particular solutions in optimal zero-sum game play. This study finds a strategy to be used after the mistake carries out in play. There are two approaches proposed to get optimal strategies. The first approach arrived from perturbed games. The second approach established on the basis of the lexicographic application. If the opponent ignores mistakes, the strategy will remain optimal as it does not turn to give a loss.

Gawlitza et al. (2012) have proposed two strategy improvement algorithms for static program analysis. One is max-strategy and the other is min-strategy for static program analysis. These algorithms perform within a common general framework to solve v-cam cave equations.

### Rock paper scissors

Rock paper scissors (RPS) is a cyclic game with three strategies. See Fig. 3. The game is 2-players zero-sum. The game rules are rock wins over scissors, scissors over paper and paper over rock. Following papers considers RPS game theoretic model.

Sinervo and Lively (1996) have used cyclic RPS game in a biological study. By using this zero-sum model they studied three different strategies of male side-blotch lizards. It studies territory use and patterns of sexual selection on male side-blotch lizards.

Bahel and Haller (2013) have computed Nash equilibria of cyclic RPS game. They characterized Nash equilibria into two sets. With an even number of actions, an infinity of Nash equilibria exists. On the second set with an odd number of actions unique Nash equilibria is found. This paper studies the strength of Nash equilibria.

Frey et al. (2013) have studied complex dynamics in social and economic systems. This is realized by analyzing agents independently playing a multiplayer mod game. The game is like the rock paper scissors. The behavior of players in human groups is non-fluctuating and effective. In this game the periodic behavior is stable.

Batt (1999) has also studied the model of Rock Paper Scissors game and has presented insights of the game having an efficient outcome with few conflicts. The game players are biased for being a winner. This game is not efficient with major conflicts. For that other approaches like coin-flip is the best choice.

Neumann and Schuster (2007) have used a zero-sum rock scissor paper game as a framework. By which they modeled the process of bacteriocin producing bacteria. The game is examined for three strains. These are of *E. coli*, bacteriocin producer, resistant and sensitive. They derived stability criteria for these strains. The paper actually proposes Lotka–Volterra system model of the RPS game.

Duersch et al. (2012) have obtained Nash equilibrium for the 2-player symmetric game. There is no pure equilibrium exists in RPS game. They found that pure equilibrium strategy exists only in non-generalized rock paper scissors game. It also showed that pure equilibrium exists for the 2-player finite symmetric game.

### Cake cutting

Cake cutting is a simple child game. See Fig. 4. In this game, the first player has to cut the cake and then the second player has to choose the piece. The first player has to cut pieces

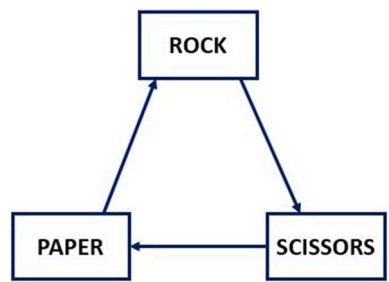

**Fig. 3** Rock paper scissors is a three strategic 2-player game. According to game rules rock beats scissors, scissors beat paper and paper beats rock. The game will draw if both players show same options

equally. Otherwise, the second player has the choice to choose either the bigger piece or the smaller one. This is to accomplish honesty in the game.

Procaccia (2013) have discussed cake cutting game. They described that it is a powerful tool to divide heterogeneous goods and resources. Cake cutting algorithm looks for formal fairness in the division of heterogeneous divisible goods. But the design of these algorithms is a complex task for computer scientists.

Edmonds and Pruhs (2006) have proposed a randomized algorithm that considers cake cutting algorithm. It equally allocates resources between n numbers of players. This algorithm needs honesty of players.

### Matching penny

Matching penny is also a zero-sum 2-player game. Both players secretly turn their coins and then compare with each other. If both are heads or tails then the first player will win else player 2 will win both coins. See Fig. 5.

McCabe et al. (2000) have studied three-person matching pennies game. It examines knowledge of player about other player's payoffs and actions. The Naive Bayesian learning and sophisticated Bayesian learning are studied in this context. These approaches examine that estimated mixed strategies can be played or not. Results showed that players do not use sophisticated Bayesian learning to obtain Nash equilibrium.

Stein et al. (2010) have studied mixed extension of matching pennies, a zero-sum game. This study constructs examples to support polynomial games. Here Nash equilibria are representable as finitely moments. Whereas polynomial games cannot be represented as finitely moments.

### Colonel Blotto

Colonel Blotto is a universal game providing a way for resource allocation. See Fig. 6. The two colonels simultaneously distribute resources over battlefields. The player devoting the most resources wins that battlefield. The payoff is equal to the total number of battlefields won.

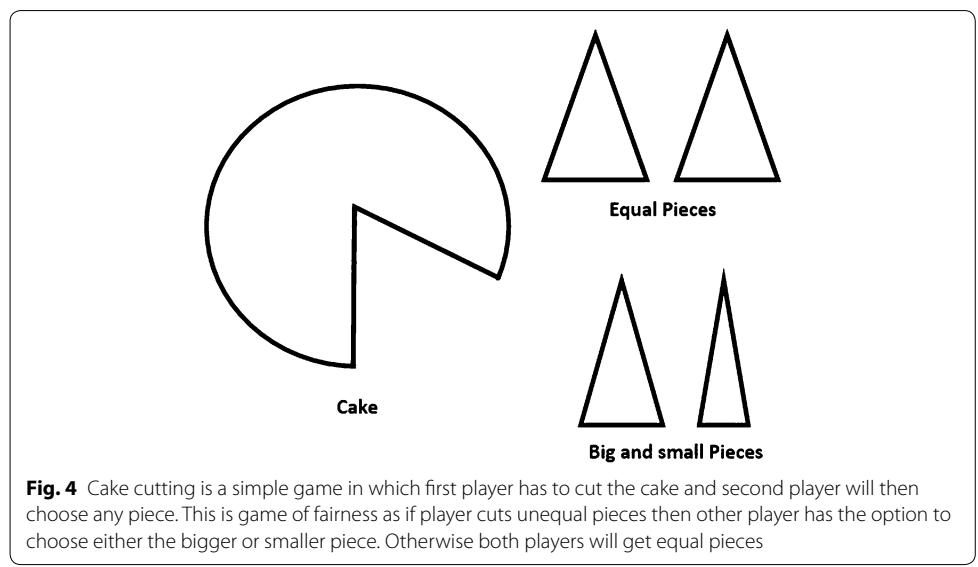

**Fig. 4** Cake cutting is a simple game in which first player has to cut the cake and second player will then choose any piece. This is game of fairness as if player cuts unequal pieces then other player has the option to choose either the bigger or smaller piece. Otherwise both players will get equal pieces

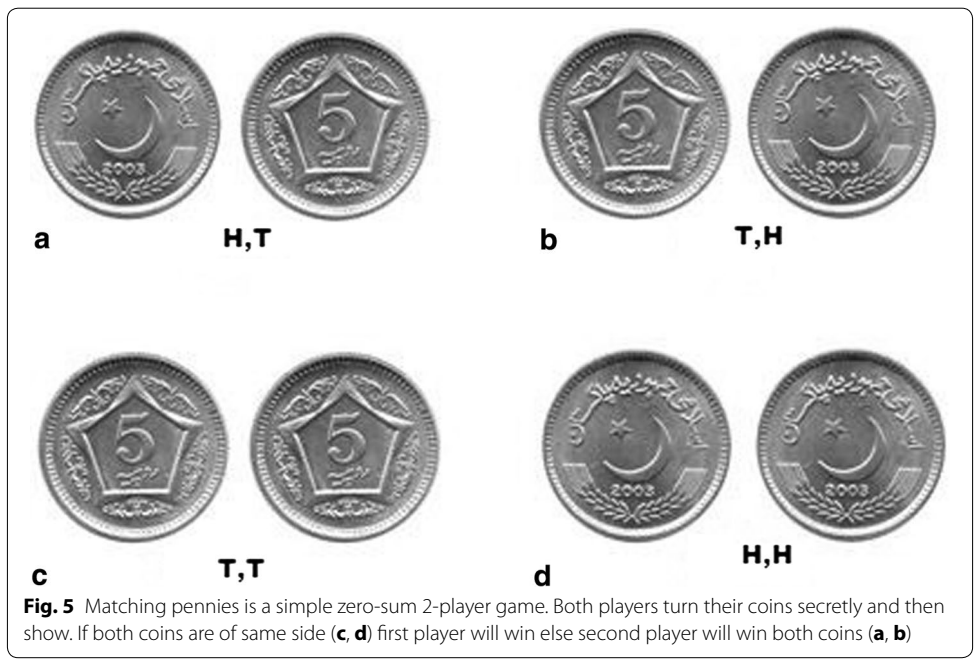

**Fig. 5** Matching pennies is a simple zero-sum 2-player game. Both players turn their coins secretly and then show. If both coins are of same side (**c**, **d**) first player will win else second player will win both coins (**a**, **b**)

Roberson ([2006](#)) described the remarkable equilibrium payoffs in the Colonel Blotto game. It considers both symmetric and asymmetric cases of the zero-sum game. The proportion of won battlefields is the payoff of player.

Hart ([2008](#)) have studied Discrete Colonel Blotto game. This is a Zero-sum game with the symmetric case for which optimal strategy is obtained. Both of these games deal with the conflicting environment.

### Kuhn Poker

Kuhn Poker is a simplified form of Poker developed by Harold W. Kuhn (Tucker [1959](#)). In this 2-player game, the deck includes only three cards. One card is distributed to each player. The first player has to bet or pass then the second player may bet or pass. On a bet, the next player must bet also. When both players pass or bet then the player with the highest card will win the pot.

Southey et al. ([2009](#)) have studied Kuhn Poker game. There main concern is opponent modeling in the game. They studied two algorithms, expert and parameter estimation. Their experiment showed that learning methods do not give good results in the small game.

### Princess Monster

Rufus Isaac formulated a game Princess Monster in his book "Differential Games" (Isaaks [1952](#)). This is a Zero-sum game between two players, Princess and Monster. The game played on 2-D search set. See Fig. [7](#). When the distance between both players is less than r then Princess got captured and Monster wins.

Wilson ([1972](#)) has developed this game on a circle. Princess and Monster move on a circle either clockwise or anti-clockwise. If both players move in the same direction, the game state does not change. But if they move in opposite directions then there will be a point on the circle on which both reach at the same time. At that point, Princess got captured and Monster wins.

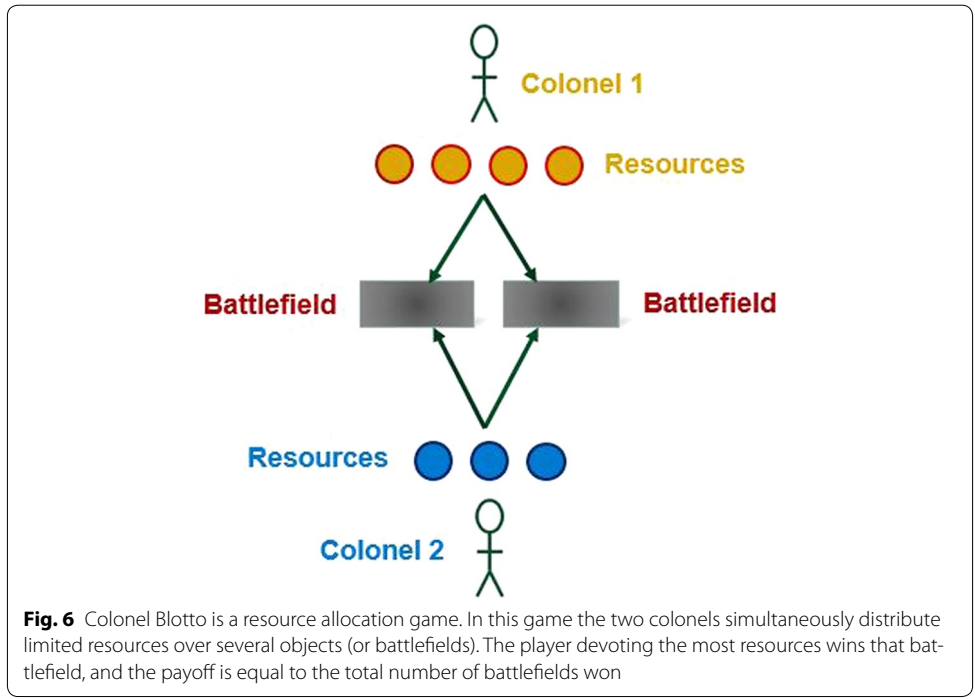

**Fig. 6** Colonel Blotto is a resource allocation game. In this game the two colonels simultaneously distribute limited resources over several objects (or battlefields). The player devoting the most resources wins that battlefield, and the payoff is equal to the total number of battlefields won

### Solution concepts

We have discussed before that game describes strategic interactions. In game theory, the solution concept is like a rule by which game theorists seeks how the game will be played. The Nash equilibrium, Pareto optimality, and Shapley values are different known solution concepts. These concepts are used to formally predict that how the game will be played.

#### Nash equilibrium of games

Nash (1951) defined Nash equilibrium. In the Nash equilibrium, all players know each other's equilibrium strategy. And no utility a player can have by changing its own strategy only. For example, there is a game battle of sexes (Shah et al. 2012). The game is between husband and wife. Husband prefers to go for football match and wife wants to go for a concert. Also, they want to go together. The payoff table is shown in Table 2. The solution for the game can be either both go for a football match or go to a concert.

Singh and Hemachandra (2014) have studied Nash equilibrium for stochastic games with independent state processes. This study got basis from the work of Altman et al. 2008. They worked on N-player Constrained Stochastic games.

Grauberger and Kimms (2014) have computed Nash equilibria for network revenue management games. This study investigates network management competition. A heuristic is presented for computing Optimal Capacity allocations. It also computes Nash equilibria in non-zero-sum games. It computes approximate to exact Nash equilibrium. They used the linear continuous model to reduced computational time.

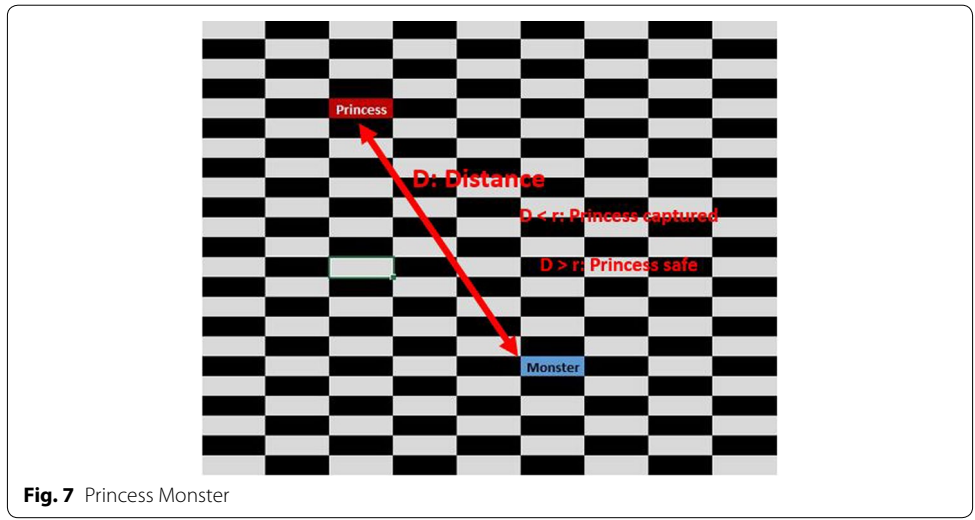

**Fig. 7** Princess Monster

Gharesifard and Cortes (2013) have considered a network based scenario and obtained a Nash solution. Network's aim is to maximize or minimize a common objective function. The two players are two network agents. They have their objectives to achieve network's aim. Both agents with opposite aims make a zero-sum game between them. Each network's saddle-point dynamics implemented by both network's through local interactions. The saddle-point dynamics for concave-convex class converges to Nash equilibrium. This saddle-point dynamics do not work to converge directed networks.

Porter et al. (2008) have proposed two search methods that calculate Nash equilibrium. One method is for the two-player game and the second method is for the n-player game. Both methods uses backtracking approaches to search the space of small and balanced support. These methods are tested on different games. Results showed positive performance of these methods. Another approach the Lemke–Houson algorithm for two-player games also discussed here.

Rosenthal (1974) have obtained correlated equilibria for 2-player games. These are more general strategies than Nash equilibrium known as correlated equilibrium. There can be a player who prefers correlated equilibria on Nash equilibrium. If this so, then correlated equilibria is a convenient solution. If the game is the best response then the correlated equilibria are not the right solution. It is good for the competitive games.

Hu and Wellman (2003) have computed Nash equilibrium for the general-sum stochastic game. They proposed a method for a multiagent Q-learning. The method Nash-Q generalizes Q-learning of single-agent to the multiagent environment. It updates its Q-function by assuming Nash equilibrium actions as a choice of agents. It is shown that Nash Q provides efficiency to get equilibrium on single-agent Q-learning. This is an offline learning process. The online version of this learning process is also implemented.

**Table 2  Payoff table of battle of sexes (Adapted from Shah et al. (2012))**

| Husband | Wife | |
|---|---|---|
| | Football | Music |
| Footbal | (3,1) | (0,0) |
| Music | (0,0) | (1,3) |

Maeda (2003) have considered games that have fuzzy payoffs. They first characterize equilibrium strategies as Nash equilibrium strategies. Then they examine characteristics of game values of fuzzy matrix games. Finally, they demonstrated this approach via numerical example.

Athey (2001) have studied games known as games of incomplete information. They proposed a restriction called single crossing condition (SCC) for these games. The Pure Strategy Nash equilibrium with a finite set of actions exists if SCC is satisfied. In these games, players have private information of their own. The results of this study show non-decreasing Pure Strategy Nash equilibrium. The proposed approach is constructive. So that the equilibria can be calculated for finite action games easily.

### Pareto optimality

Pareto optimality introduced by Vilfredo Pareto (Yeung 2006). In Pareto optimal game, there exists a strategy that increases player's gain without damaging others. For example, when Economy is competitive perfectly then it is Pareto optimal. This is because no changes in the Economy can make better the gain of one person and can make worse the gain of another person at the same time.

Feldman (1973) has discussed Pareto Optimality in bilateral barter. The proved the constraints under which trade moves go on to pairwise optimal allocation. Then this paper discussed some general conditions by which these allocations are Pareto optimal.

Kacem et al. (2002) have solved the flexible job-shop scheduling problem.by using hybrid Pareto approach. Their proposed approach combines Fuzzy logic and evolutionary algorithms. This combination minimizes machine workloads and completion time.

Guesnerie (1975) have discussed insights of non-convex economics. The paper characterizes Pareto-optimal states. Then analyze how to achieve them in distributed economy. The focus of this paper mainly concerns with conditions needed for optimality, marginal cost pricing rules, and decentralized non-convex economy.

### Shapley values

There is a Shapley value another solution concept used in cooperative game theory (Shapely 1953). It allocates a distribution to all players in a game. The distribution is unique and the game value depends on some desirable abstract characteristics. In simple words, Shapley value assigns credit among a group of cooperating players. For example, there are three red, blue and green players. The red player cooperates more than blue and green players. The goal is to form a pair and then assign credits to them. Each pair must have a red player as it cooperates more than others. So there can be two possible pairs. The two pairs are:

1. Red player, blue player
2. Red player, green player.

The red player cooperates more, so it will get more profit than player blue in the first pair. Similarly, it will get more profit than a green player in the second pair.

Littlechild and Owen (1973) discussed the problem of computing Shapley value for large games. They considered the work of Broker and Thompson of about aircraft landing charges on the airport. This paper presents an expression that can be calculated when the cost function is a characteristics function. The costs of the biggest player in any subset of players is equal to the cost of that subset.

Gul (1989) has worked on the bargaining problem in a transferable utility economy. A framework is established by which the two approaches, cooperative and noncooperative, are compared. The stationary subgame perfect Nash equilibrium is used and with small time intervals, the gain is the Shapley value for the agent.

Pérez-Castrillo and Wettstein (2001) have proposed a mechanism to analyze how cooperation produces surplus. It is a two-phased play. The first phase is of bidding that gives the winner of the game. In the second phase the winner is rejected then the game is again played without that winner. This paper describes that the payoff of the game coexists with Shapley value.

### Decision theory

Parsons and Wooldridge (2002) have discussed both game and decision theories. As game theory studies agent's interaction, it is closed relative to decision theory. Decision theory seeks to get the most favorable choice. That can maximize utilities of decision makers. Whereas the game theory also studies self-interested agents. It takes agents as greedy players want to maximize their own gain. This paper reviewed existing literature. Then it revealed issues related to autonomous agents and multi-agent system.

Hart et al. (1994) have worked on the two-person zero-sum game. They obtained game value and derived utility simultaneously by using decision theory. They found the gap between the axioms and presumption about expected utility maximization. Axioms characterize expected utility maximization, considering risk, in the individual decision. The presumption is that expected utility maximizers evaluate the game by their value. This study does not fill this gap completely. Because rationality involves playing maximin strategies is not proved.

### Game theory in computer science

Roughgarden (2010) have described Algorithmic Game Theory (AGT), a game theory applications in computer science. This paper explores current research formats in AGT. The research theme is different here than classical game theory. AGT receives the computational difficulty as a coupling requirement which makes it unique.

Wooldridge (2012) have explored the feasibility of game theory applications in computer science. They discussed issues related to the application of game theoretic models. They revealed the incorrect use of game theory model. They also mentioned that more research is needed in this area.

Ahmad and Luo (2006) have proposed an algorithm for video coding. It considers optimization of rate control. In this two-level algorithm, the first level is about the target bits allocation. In the second level, each MB computes to share bits fairly. So that its quantization scale can be optimized.

**Games in social systems**

In this section, we will discuss game theory applications in social groups and others. In social groups, people interact and communicate each other. To model behaviors in such communication, game theory has been used.

Chen and Liu (2012) have modeled human behavior in social networks by using game theory. This is the study of the impact of social networks in our daily life. This generalized approach can be used for several social networks. The efficiency and fairness between users are main considerations of the model design.

Hand (1986) has discussed social conflicts and social dominance. The social dominance based on Leverage is considered here. There are personals having greater resources and personals having fewer resources as well. The paper describes that game theory can be used to make less dominant individuals equal or greater to others.

**Markov games**

Altman (1994) have used Markov games to control the flow of arriving packets. These are the collection of normal-form games that agents play repeatedly. These games together with a value iteration algorithm are used for single controller. The controller design policies to control the flow. Markov games is another name of stochastic games. This study reveals the existence of the stationary optimal policy.

Ghosh and Goswami (2008) have studied semi-Markov game. They first transformed the model into the completely observed semi-Markov game. Then they worked and obtained saddle-point. They showed the existence of saddle-point but with some conditions.

Laraki et al. (2013) have discussed stochastic games, subgame perfect and Borel sets. It describes conditions for the existence of game value. With these conditions the player 2 gets an optimal strategy for subgame perfect. The conditions described that payoff is a bounded function f. The function f is measurable and is lower semi-continuous.

Deshmukh and Winston (1978) have developed zero-sum model for product's price setting in two firms. The model is based on some assumptions. That is the current price of product and market positions influenced future market positions. This provides a way to get balance benefits gained from price variations.

Sirbu (2014) has studied zero-sum games. The paper discussed stochastic differential game restricted to elementary strategies. The result shows the existence of value in a game with these strategies.

Pham and Zhang (2014) have studied 2-player zero-sum weak formulation game. The game discussed is Stochastic and Differential game. The game value is obtained by visocsity solution. The paper showed the value of the game as a random process.

Hernandez-Hernandez et al. (2015) have studied Stochastic Differential Equation. The game is between controller called minimizer and stopper called maximizer. The controller selects a finite-variation process. And the stopper selects time at which the game will stop. The study described that the obtained optimal strategies are not unique.

Oliu-Barton (2014) has worked on Finite Stochastic game. This is a zero-sum game. The paper proves the presence of value in the game. The aim of the study is to provide asymptotic behavior of strategies.

Hamadène and Wang (2009) have studied Backward Stochastic Differential Equations. These equations have terms. Their resulted solution is also a stochastic or random process. The paper presents a remarkable solution and showed the value in the game.

Shmaya (2006) have studied an interesting game with one informed player. It is a two-player zero-sum game with stochastic signals. The value of the game is taken as a function of player one's information structure. The properties of this function, examined, shows that every player has a positive value of information in zero-sum game.

### Non-zero-sum game models

In non-zero-sum games, there exists a universally agreed solution. It means there is no single optimal solution as zero-sum games have. These games model cooperation instead of conflicts. There can be a win-win solution of game where everyone is a winner. The players can play a game while cooperating each other to achieve a common goal.

Sullivan and Purushotham (2011) have discussed a high-level summit on non-communicable disease (NCD). The summit held in New York on September 2011 in which they discussed cancer policies. The summit recognized cancer a first high-level disease. This paper critically examined these policies. It gives an alternative solution based on a non-zero-sum game model for international cancer policy.

Bensoussan et al. (2014) have worked on the non-zero-sum stochastic differential game. They modeled performance of two insurance companies. Each company is greedy to maximize its own utility. The surplus process modeled by a continuous-time Markov chain and an independent market-index process. The game solved by a dynamic programming principle. It is also mentioned that the presented game can be extended to several directions.

Carlson and Wilson (2004) have considered failure in the management of U.S. national forest. At first, this seems a pure conflict between US National Forest Service and Environmentalists. But in this paper, a non-zero-sum game theoretical model is developed. It examines the effects of these changes on outcomes. It is analyzed that some changes do not affect outcomes and some have potential impact.

Shenoy and Yu (1981) have studied partial conflict games. This study examines the reciprocative strategy to induce cooperation. Reciprocative behavior is defined as Non-Zero-sum games. It describes conditions for cooperative behavior to give an optimal response to reciprocative behavior. The feasibility of playing reciprocative strategy is also determined. Finally, conditions are given for reciprocative strategy that results to Nash equilibrium.

Mussa (2002) have studied two monetary units, euro, and dollar. This article argues that there is a non-zero-sum game between both units. It defines euro beneficial for both the euro area itself and rest of the world. Euro effects world's economy indirectly. It is described that euro and the dollar are co-equal monetary standards. And is beneficial to the United States, euro area itself and rest of the world.

Semsar-Kazerooni and Khorasani (2009) have studied multi-agent system that considers cooperative game theory. The common goal of the multi-agent team is to have consensus. Consensus can be accomplished over a common value for the agent's output. This paper is a series of work. In this paper, a previously introduced strategy is used called semi-decentralized optimal control strategy.

Khosravifar et al. (2013) have used an agent-based game theoretic model to analyze web services. There is a distributed environment in which agent cooperates each other. The performance of agents is analyzed by using non-zero-sum model. The decision-making process is also analyzed.

Radzik (1991) have obtained pure-strategy and Nash equilibrium for 2-player non-zero-sum games. The payoff functions are upper semicontinuous. Agents are not allowed to interact each other in the model considers here. The optimality criterion dominant is the NE vector. This vector computes optimal actions of all players considering their payoff function. The paper emphasizes solutions in pure strategies.

Radzik (1993) have computed Nash equilibria for discontinuous two-person non-zero-sum games. This study examines two classes of these games on the unit square. Here the payoff function of the first player is convex or concave in the first variable. This supposition combined with bounded payoff function entail the presence of Nash equilibria.

### Games in networks

The networks provide an excellent way of communication as well as support for distributed environments. The Game theory models have their obvious applications in network-based systems. The following papers use game theory to get optimal strategies for network problems.

#### Transport networks

Bell et al. (2014) have proposed a game theoretic approach for modeling degradable transport networks. By this approach, hyperpaths are generated between population centers and depot locations. They used a case study in the province of China to facilitate the proposal. Optimal hyperpaths are defined by using mixed strategy Nash equilibrium. Which give ultimate depot locations. These depot locations are found by using two forms of drop heuristic. These heuristics gives optimal solution except in one case. That is when the most appropriate location for only one rescue center is obtained.

Alpcan and Buchegger (2011) have studied vehicular networks. They examine security of network for the improvement of transportation. It is to provide optimal strategies to defend malicious threats. Three types of security games are studied here. When players knows the payoff matrices the game is a zero-sum. When they know approximate payoffs the game is a fuzzy game. When players do not know each other's payoffs, strategies can be improved via fictitious play.

#### Network security and reliability

Perea and Puerto (2013) have used game theory approach in network security. The game is between the network operator and attacker. The operator establishes network to achieve some goals. While the attacker wants to place damages in the network. The optimal strategy for the operator is building a network. The optimal strategy for attacker is finding edges to be attacked. This paper revealed dynamic aspects of the game.

Bell (2003) has proposed a novel method to identify failure nodes. It is a two-player game between a router and virtual network tester. Router has to find a least-cost path, whereas network tester wants to increase trip-cost. The link in use are optimal for router and failure links are optimal for network tester. Network tester fails link to increase trip-cost. So the given maximin method is to identify those links that threaten to network.

Kashyap et al. (2004) have modeled multiple-input/output fading channel communication problem as a Zero-sum game. The players, maximizer and minimizer, have mutual information. On both maximizer and minimizer there is total power constraint. They obtained saddle-point of the game. It is shown that minimizer has no need of channel input knowledge.

Wei et al. (2012) have applied game theoretic approach for a non-correlated jamming problem. In this problem jammer has a lack of information about actually transmitted signals. There is a Zero-sum game between transceiver pair and jammer in the parallel fading channel. This paper explored CSI and solved problems related to it. The study finds equilibrium based on pure strategy. The game model adopts frequency hopping to defend against jam threats.

Chen et al. (2013) have used the zero-sum game model to analyze the performance of system. The approach examines communication across cooperative and malicious relays. It also analyzes the impact of this communication. The malicious relays can jam the network and they intentionally interrupt the system. The Nash equilibrium is determined to get optimal signaling strategies for cooperative relays.

Venkitasubramaniam and Tong (2012) have studied network communication. They used zero-sum game theoretic approach to provide anonymity. Optimizing anonymity problem is a game between network designer and adversary. The model showed the presence of saddle-point. The approach obtained optimal strategies by using parallel Relay networks. It explores throughput tradeoffs in large networks.

Wang and Georgios (2008) have considered Jammer and Relay problem. They modeled the problem between them as zero-sum mutual information game. By assuming source and destination being unaware optimal strategies are derived for both players. In non-fading scenario Linear Relay (LR) and Linear Jammer (LJ) are optimal strategies. In fading scenario, J cannot distinguish between Jamming and source signal. So the best strategy is to jam with Gaussian noise only. Here R forward with full power when jam link is worst. They derived optimal parameters on the basis of exact Nash equilibrium.

Zhao et al. (2008) have studied Wireless Mesh Networks. They used game theoretic approach for increasing performance of MAC protocols. This is an iterative game having two steps. In the first step current state of the game is determined on each node. In a second step, the equilibrium strategy of the node is adjusted to the determined state of the game. The process is repeated till the desired performance is achieved. Finally, results are validated via simulation.

Larsson et al. (2009) have studied signal processing and communications in a game theoretic way. They demonstrated basic concepts of conflicting and cooperative game theory through three examples of interference channel model. These are SISO IFC, MISO IFC, and MIMO IFC. For conflicting case the study is limited to Nash equilibrium and price of anarchy (PoA). The Price of anarchy gives cost measures that system paid to operate without cooperation.

Nguyen et al. (2013) have used game theory to integrate distributed agent-based functions. They proposed an agent-based conceptual strategy. Which resolves the conflicting interests between product agents and network agents. The method is based on cooperative game theory that integrates and solves conflicting interests. Finally, the approach is verified by simulation with two case studies. First is like micro grid example and the second is the more complex case.

Quer et al. (2013) have used game theoretic approach to study inter-network cooperation. The scenario is about two ad hoc wireless networks. Both cooperates together to gain some benefits. Statistical correlation between local parameters and performance is computed by Bayesian networks method. Both networks share their nodes to achieve cooperation. Game theory is used in nodes selection process. The system level simulator is used to confirm results. Results showed that increase in performance can be achieved by accurate selection of nodes.

Spyridopoulos (2013) have modeled problem of cyber-attacks. For that, they used Zero-sum one-shot game theoretic model. Single-shot games are opposed to repeated games. These models can be used when cooperation cannot be possible among players. The study explored adjustments and ideal techniques for both assailant and keeper. The study revealed a solitary ideal method for the keeper. The ns2 network simulator is used for the simulation of the model.

Khouzani et al. (2012) have studied software-based operations against malware attackers. Malware has to maximize the damage. And the network has to take robust defensive strategies against attacks. This makes the game a Zero-sum game. Simple robust defensive strategies are shown via dynamic game formulation. Finally, performance is evaluated through simulation.

### Discrete-time/continuous-time

Ye et al. (2013) have proposed a discrete-time Markov chain Parrondo's model. They analyzed model theoretically and verified via simulation. One can realize rationality and adaptability from a macro level. They showed that agitating effect of rewiring is effective than the zero-sum game.

Al-Tamimi et al. (2007) have proposed an algorithm for the solution of a zero-sum game. The algorithm provides a solution for Riccati equation. They discussed two schemes of programming. One is heuristic dynamic and second is dual. These schemes used for the solution of the value function and game costate.

Liu et al. (2013) have proposed an algorithm based on finding approximate optimal controller. It is based on the class of discrete-time constrained systems. This iterative adaptive dynamic programming algorithm provides a solution for near-optimal control problem. The control scheme has three neural networks. These networks are taken as parametric structures to assist the proposed algorithm. This is described by two examples that showed the practicality and concurrence of the algorithm.

Wu and Luo (2013) have modeled H∞ state feedback control problem as the two-person Zero-sum game. An algorithm is proposed for solving algebra rectaii equation. They developed two versions, offline and online. An offline version is a model-based approach. The online version is a model-free approach but partially. These approaches are validated through simulation.

Abu-Khalaf et al. (2008) have used policy iteration approach together with neural networks. They provide practical solution method for suboptimal control of constrained input systems. They modeled the problem as a continuous-time zero-sum game. The study showed new results and creates a least-squares-based algorithm for a practical solution. The proposed algorithm is applied to the RTAC nonlinear benchmark problem.

### Resource allocation

Zhou et al. (2011) have modeled energy allocation problem in two phased training-based transmission. The model is based on the zero-sum game between two phases. The two phases are training phase and transmission phase. This study is about optimal energy allocation between these two phases. The closed-form solutions are derived from jammer's view. The study proves the presence of NE for fixed training length. Finally, it discusses channel state information.

Tan et al. (2011) have discussed radio networks. They used game theory approach for fair sub-carriers allocation and power allocation. The sub-carrier allocation and power allocation are based on colonel blotto game. The secondary users allocate budget wisely to transmit power to win sub-carriers. Power allocation and budget allocation are strategies used for fair sharing among secondary users. This paper proposed algorithms and conditions for the presence of unique NE. Finally, the results are validated through simulation.

Belmega et al. (2009) have discussed power allocation in fast fading multiple access channels. In these channels transmitters and receiver have many antennas. The study gives unique Nash equilibrium. It also gives best power allocation policies. The paper discussed two different games. In the first game, the users can adapt their temporal power allocation to their decoding rank at the receiver. The other is to optimize their spatial power allocation between their transmit antennas. Finally, results are shown via simulation.

In the next section, we will classify games in tabular structures. Then will discuss some open problems.

## Discussion

We discussed game theory and its applications in different domains by exploring different papers. We described how game theory models strategic and complex interactions of self-interested agents. We also proposed a general taxonomy of games, based on the types of game representation. The three types of game representation are Normal-form, Extensive-form, and Beyond Normal/Extensive form. Then we classify games according to these representation types.

We have seen different games while reviewing literature. Such as Markov games, Zero-sum game, Stochastic game, Bayesian games etc. These are actually different classes of games having different properties. We summarized different games, by their different types. See Table 3. The legend used in the table is summarized in Table 4.

We also summarized games discussed in different papers according to representation forms. The representation forms are Normal, Extensive and Beyond normal/extensive form. See Table 5.

## Open problems

We have noted that while researchers applied game theory in different domains, there is still need to further exploit game theory in the modeling of complex systems research. In computer science, there is also a need to apply game theory in the domain of resource allocation algorithms such as in clouds, Internet of Things, Cyber physical systems, and others. Cake Cutting and Colonel Blotto are quite possibly good game-theoretic resource allocation models and can thus be used in such domains. However, they have not previously been used much in these areas. Furthermore, fair allocation is still a complex task in distributed systems. With the advent of mobile, pervasive computing, and cloud-based systems, practical distributed computing requires the resolution of such dilemmas on a regular basis. In other words, there is a growing need to use game theory for practical applications in the technological domains rather than restrict it to purely theoretical applications and those too, limited to very specific and niche areas of research.

Another open area for further research is in the development of taxonomies for specific game theoretic areas. We have proposed a general taxonomy of games. We have also mentioned few previously defined taxonomies. However, there is a need for the development of more taxonomies of games. These include the development of taxonomies and review of papers and games such as in the domain of Bayesian games, Congestion games among others.

## Conclusions and future work

This paper presents a review of game theory models from the agent-based modeling perspective. We have discussed different classes of games such as Zero-sum, Perfect information, Bayesian, Congestion etc. We have also explored the importance and nature of game theory by means of a novel taxonomy. The presented taxonomy of game classes has been based on types of game representation. In the review, game theory applications in different fields has also been discussed. We believe that this review will help multidisciplinary researchers in expanding their knowledge about the state-of-the-art in game theory. In particular, it will help researchers to look at game-theoretic literature analyzed from the perspective of agents and complexity.

**Table 3 This table lists games defined in different publications**

| Games | References | Forms | Zero-sum | Perfect | Stochastic | Repeated | Bayesian | Congestion |
|---|---|---|---|---|---|---|---|---|
| RPS | | | | | | | | |
| Three-morph mating | Sinervo and Lively (1996) | N | Z | I | No | No | No | No |
| Extended RPS | Bahel and Haller (2013) | N | Z | I | No | No | No | No |
| Mod game | Frey et al. (2013) | B | NZ | I | No | Yes | No | No |
| Continuous RPS | Neumann and Schuster (2007) | N | Z | I | No | No | No | No |
| Cake cutting | | | | | | | | |
| Balls and bins | Edmonds and Pruhs (2006) | B | Z | P | No | Yes | No | No |
| Matching pennies | | | | | | | | |
| 3-player MP | McCabe et al. (2000) | B | Z | P/I | No | Yes | Yes | No |
| Blotto games | | | | | | | | |
| Colonel Blotto | Roberson (2006) | B | Z | I | No | No | No | Yes |
| Discrete Colonel Blotto | Hart (2008) | B | Z | I | No | Yes | No | No |
| Princess Monster | | | | | | | | |
| PM on circle | Wilson (1972) | B | Z | I | No | Yes | No | No |
| Poker | | | | | | | | |
| Kuhn Poker | Southey et al. (2009) | E | Z | I | No | No | No | No |
| Networks | | | | | | | | |
| Flow control | Altman (1994) | B | Z | P | Yes | Yes | No | No |
| Network revenue | Grauberger and Kimms (2014) | B | NZ | P | Yes | Yes | No | No |
| Railway network | Perea and Puerto (2013) | N | Z | P | No | No | No | No |
| VANET security model | Alpcan and Buchegger (2011) | B | Z | P | No | Yes | No | No |
| Anonymous networking | Venkitasubramaniam and Tong (2012) | B | Z | I | No | Yes | No | No |

**Table 3 continued**

| Games | References | Forms | Zero-sum | Perfect | Stochastic | Repeated | Bayesian | Congestion |
|---|---|---|---|---|---|---|---|---|
| Jammer-relay | Wang and Georgios (2008) | B | Z | P/I | No | Yes | No | No |
| Dynamic game | Khouzani et al. (2012) | B | Z | P | Yes | Yes | No | No |
| Parrondo's model | | | | | | | | |
| Link A + game B | Ye et al. (2013) | E | Z | P | Yes | Yes | No | No |
| Transmission | | | | | | | | |
| E–D vs jammer | Kashyap et al. (2004) | B | Z | I | No | Yes | No | No |
| Transmission security | Chen et al. (2013) | B | Z | I | No | Yes | No | No |
| Payoff games | | | | | | | | |
| Average payoff | Ghosh and Goswami (2008) | B | Z | I | Yes | Yes | No | No |
| Semicontinuous payoff | Laraki et al. (2013) | B | Z | I | Yes | Yes | No | Yes |
| Symmetric | | | | | | | | |
| Symmetric game | Duersch et al. (2012) | N | Z | I | No | No | No | No |
| Mixed zero-sum | | | | | | | | |
| Mixed-strategy | Seo and Lee (2007) | B | Z | P | No | Yes | No | No |
| Mixed zero-sum | Hamadène and Wang (2009) | B | Z | I | Yes | Yes | No | No |
| Searching | | | | | | | | |
| AGTCS2-player search | Zoroa et al. (2012) | B | Z | P | No | Yes | No | No |
| Investments | | | | | | | | |
| Insurance games | Bensoussan et al. (2014) | B | NZ | P | Yes | Yes | No | No |
| Duopoly | | | | | | | | |
| Duopoly game | Deshmukh and Winston (1978) | B | Z | I | Yes | Yes | No | No |
| Others | | | | | | | | |
| Web services | Khosravifar et al. (2013) | B | NZ | I | No | No | No | Yes |

**Table 4  Legends used in Table 3**

| Legends | Name |
| --- | --- |
| N | Normal-form |
| E | Extensive-form |
| B | Beyond normal/extensive |
| Z | Zero-sum |
| NZ | Non-zero-sum |
| P | Perfect |
| I | Imperfect |

**Table 5  Games in different forms of representation**

| S. no | Ref | Games | Normal | Extensive | Beyond N/E |
| --- | --- | --- | --- | --- | --- |
| 1 | Three-morph mating | Sinervo and Lively (1996) | Yes | No | No |
| 2 | Extended RPS | Bahel and Haller (2013) | Yes | No | No |
| 3 | Mod game | Frey et al. (2013) | No | No | Yes |
| 4 | Continuous RPS | Neumann and Schuster (2007) | Yes | No | No |
| 5 | Balls and bins | Edmonds and Pruhs (2006) | No | No | Yes |
| 6 | 3-player MP | McCabe (2000) | No | No | Yes |
| 7 | Colonel Blotto | Roberson (2006) | No | No | Yes |
| 8 | Discrete colonel Blotto | Hart (2008) | No | No | Yes |
| 9 | PM on circle | Wilson (1972) | No | No | Yes |
| 10 | Kuhn Poker | Southey et al. (2009) | No | Yes | No |
| 11 | Flow control | Altman (1994) | No | No | Yes |
| 12 | Network revenue | Grauberger and Kimms (2014) | No | No | Yes |
| 13 | Railway network | Perea and Puerto (2013) | Yes | No | No |
| 14 | VANET security model | Alpcan and Buchegger (2011) | No | No | Yes |
| 15 | Anonymous networking | Venkitasubramaniam and Tong (2012) | No | No | Yes |
| 16 | Jammer-relay | Wang and Georgios (2008) | No | No | Yes |
| 17 | Network-malware dynamic game | Khouzani et al. (2012) | No | No | Yes |
| 18 | Link A + game B | Ye et al. (2013) | No | Yes | No |
| 19 | E-D vs jammer | Kashyap et al. (2004) | No | No | Yes |
| 20 | Transmission security | Chen et al. (2013) | No | No | Yes |
| 21 | Average payoff | Ghosh and Goswami (2008) | No | No | Yes |
| 22 | Semicontinuous payoff | Laraki et al. (2013) | No | No | Yes |
| 23 | Symmetric game | Duersch et al. (2012) | Yes | No | No |
| 24 | Mixed-strategy | Seo and Lee (2007) | No | No | Yes |
| 25 | Mixed zero-sum | Hamadène and Wang (2009) | No | No | Yes |
| 26 | AGTCS2-player search | Zoroa et al. (2012) | No | No | Yes |
| 27 | Insurance games | Bensoussan et al. (2014) | No | No | Yes |
| 28 | Duopoly game | Deshmukh and Wayne (1978) | No | No | Yes |
| 29 | Web services | Khosravifar et al. (2013) | No | No | Yes |

**Authors' contributions**
AF and MN both contributed equally in the paper. Both authors read and approved the final manuscript.

**Author details**
[1] Software Engineering Department, Bahria University, Islamabad, Pakistan. [2] Computer Science Department, COMSATS Institute of IT, Islamabad, Pakistan.